\documentclass[conference]{IEEEtran}
\IEEEoverridecommandlockouts
\usepackage[utf8]{inputenc}
\usepackage{cite}
\usepackage{cuted}
\usepackage{subfigure}
\usepackage{amsmath,amssymb,amsfonts}
\usepackage{bm}
\usepackage{cancel}
\usepackage[ruled,linesnumbered]{algorithm2e}
\usepackage{algorithmic}
\usepackage{graphicx}
\usepackage{textcomp}
\usepackage{xcolor}
\usepackage[export]{adjustbox}
\usepackage{xurl}
\usepackage{comment}
\usepackage{paralist} 
\usepackage{enumitem} 
\usepackage[T1]{fontenc}
\DeclareMathAlphabet{\mathpzc}{OT1}{pzc}{m}{it}
\usepackage{tabularx}
\usepackage{booktabs}
\usepackage{multirow}
\usepackage{mathtools}
\usepackage{makecell}
\usepackage{soul}
\usepackage{balance}
\usepackage{hhline}
\usepackage{booktabs}
\usepackage{enumitem}

\usepackage{comment}

\usepackage[most]{tcolorbox}

\usepackage{nomencl}
\makenomenclature
\usepackage{etoolbox}
\renewcommand\nomgroup[1]{%
  \ifstrequal{#1}{P}{\vspace{10pt}\item[\textbf{Parameters}]}{%
  \ifstrequal{#1}{V}{\vspace{10pt}\item[\textbf{Variables}]}{}}{%
  \ifstrequal{#1}{S}{\vspace{10pt}\item[\textbf{Sets}]}{}}%
}

\title{Hurricane and Storm Surges-Induced Power System Vulnerabilities and their Socioeconomic Impact}

\author{\IEEEauthorblockN{Abodh Poudyal$^\dagger$, Shishir Lamichhane, Charlotte Wertz, Sajjad Uddin Mahmud, Anamika Dubey}
\IEEEauthorblockA{Washington State University, Pullman, Washington, USA}
Email: $^\dagger$abodh.poudyal@wsu.edu
\thanks{This work was supported by National Science Foundation (NSF) CAREER Grant \#1944142 and the PNNL-WSU Distinguished Graduate Research Program (C.W.)}
\vspace{-0.4cm}}
\begin{document}
\bstctlcite{IEEEexample:BSTcontrol}
\maketitle
\thispagestyle{plain}
\pagestyle{plain}

\begin{abstract}

This paper introduces a probabilistic framework to quantify community vulnerability towards power losses due to extreme weather events.
To analyze the impact of weather events on the power grid, the wind fields of historical hurricanes from 2000–2018 on the Texas coast are modeled using their available parameters, and probabilistic storm surge scenarios are constructed utilizing the hurricane characteristics. The vulnerability of hurricanes and storm surges is evaluated on a 2000-bus synthetic power grid model on the geographical footprint of Texas. The load losses, obtained via branch and substation outages, are then geographically represented at the county level and integrated with the publicly available Social Vulnerability Index to evaluate the Integrated Community Vulnerability Index (ICVI), which reflects the impacts of these extreme weather events on the socioeconomic and community power systems. The analysis concludes that the compounded impact of power outages due to extreme weather events can amplify the vulnerability of affected communities. Such analysis can help the system planners and operators make an informed decision.

\end{abstract}
\begin{IEEEkeywords}
Extreme weather events, power system resilience, social vulnerability, community vulnerability assessment
\end{IEEEkeywords}

\vspace{-0.55cm}
\section{Introduction} 



Over the past several years, there has been a rise in the severity and occurrence of extreme weather events such as hurricanes and floods. Hurricanes have accounted for trillions of dollars in damages, and the power grid frequently bears the majority of these losses~\cite{disaster}. The consequences of these hurricanes originate from the combination of high-intensity wind and the large-scale flooding generated by storm surges in coastal areas. For example, in 2023, hurricane Idalia, a category 3 storm, caused a power outage for more than 300,000 customers over the U.S. East Coast, from Florida to North Carolina~\cite{idalia}. Hurricane Ian incurred a total cost of 112.9 billion dollars in damages in 2022 and inflicted power outages on around 2.7 million consumers~\cite{ian}. 


Unfortunately, the communities at the forefront of these natural disasters are often the most vulnerable. Outages that are 8+ hours - often caused by these high-impact, low-probability (HILP) events - are experienced mostly by highly vulnerable communities~\cite{USA_outages}. During Hurricane Harvey in 2007, it was found that households with a lower socioeconomic status faced more extensive flooding than those of higher incomes~\cite{COLLINS}. Additionally, the Federal Emergency Management Administration (FEMA) has been criticized for having regressive policies regarding restoration allocation after major power outages~\cite{FEMA}. 

Existing studies have analyzed power system impact from hurricane-flood events~\cite{souto2022power, panteli2015influence}. There is minimal literature that combines community impact evaluation with power system impact. Energy equity metrics are still underdeveloped~\cite{Energy_Equity}, despite it being well-documented that inequities are apparent in outages from HILP events.In~\cite{ASCE} analysis regarding the infrastructure losses due to Hurricane Harvey is investigated and identifies impacted communities. However, the work is based on critical infrastructural information, affecting the extension of the work to other regions.

This framework analyzes historical hurricane tracks and their associated coastal flooding. The impact of these events on the power grid is modeled using component fragility curves of transmission line segments and substations.
We leverage a 2000-bus synthetic power grid model based on non-proprietary data but realistic grid parameters~\cite{syntheticgrid_Adam}. The probabilistic loss metric is evaluated using Monte Carlo Simulations (MCS), and the outage data is mapped to the county level to evaluate the community impact of these hurricanes, looking at both outage and social vulnerability. This study can be easily referenced or modified when investigating the impact a future HILP event can have on not only a power system but also the communities that rely on it.

\section{Extreme Weather Events and Impact Model} \label{sec:weather_impact_model}
\subsection{Hurricane Wind Field and Storm Surges}
\begin{figure}[ht]
    \centering
    \includegraphics[width=0.6\linewidth]{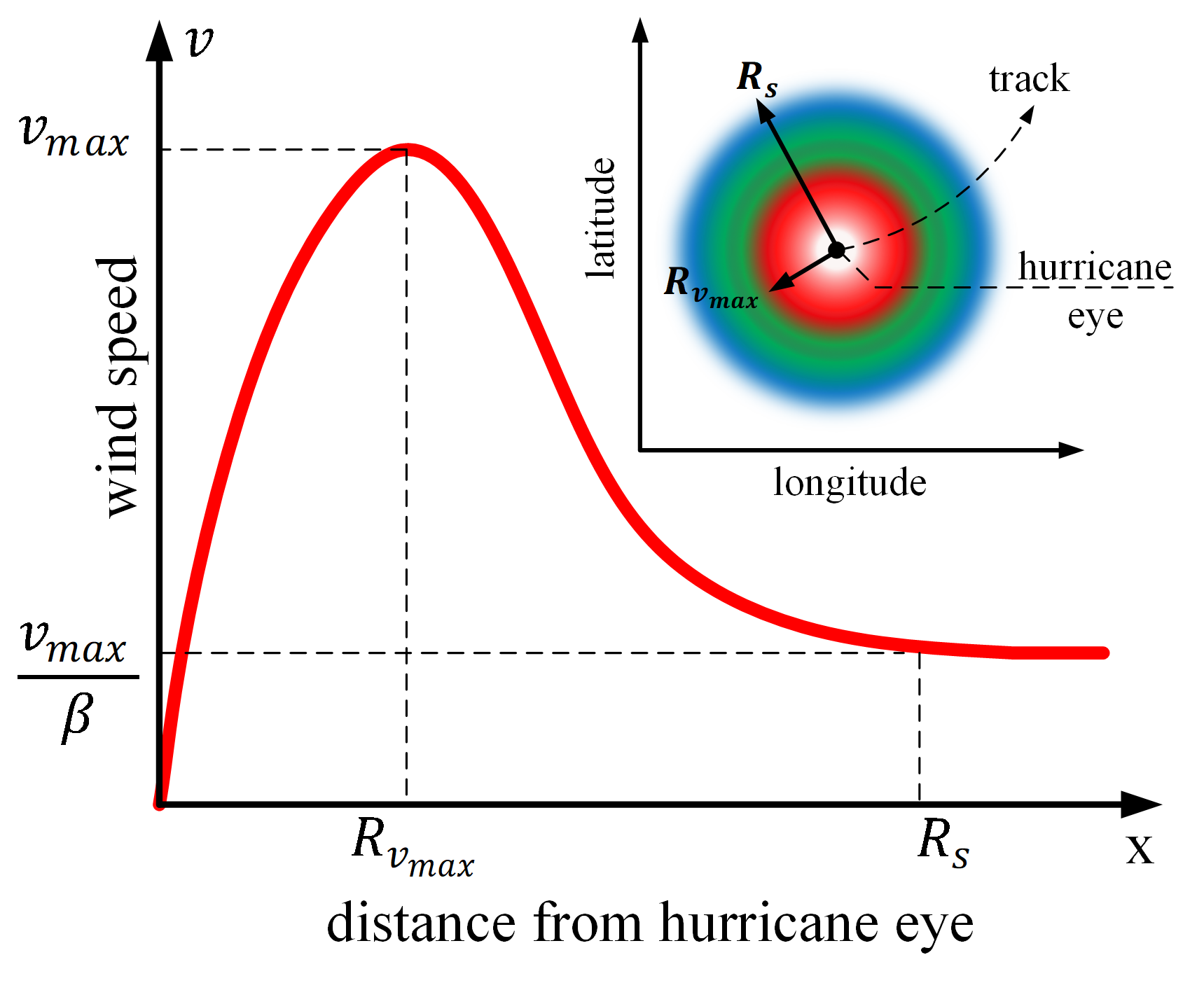}
    \vspace{-0.2cm}
    \caption{Distribution of wind field of a hurricane at a fixed time.}
    \vspace{-0.5cm}
    \label{fig:static_hurricane}
\end{figure}

We use a statistical model to determine the hurricane wind field at each time step ($t$) as described in~\cite{poudyal2023spatiotemporal}. The wind field model of a hurricane at each $t$ is dependent on three different parameters  -- namely max sustained wind speed ($v_{max}$), radius from the hurricane eye to $v_{max}$ ($R_{v_{max}})$, and radius of the hurricane ($R_s$), known as the radius of the outermost closest isobar (roci). These parameters are available from the International Best Archive for Climate Stewardship (IBTrACS) for historical hurricanes and are recorded every 3 hours~\cite{Knapp2010}. These parameters form the hurricane's wind field at each $t$. The distribution of the hurricane's wind field at any fixed time is shown in Fig.~\ref{fig:static_hurricane}. Here, $\beta$ refers to the decrement factor of $v_{max}$ at $R_s$.      

In this work we only consider coastal flood scenarios and leverage the SLOSH (Sea, Lake, and Overland Surge from Hurricanes) model developed by the National Oceanic and Atmospheric Administration (NOAA)~\cite{glahn2009role}. 
SLOSH covers the entire U.S. Atlantic and Gulf of Mexico coastlines subdivided into 32 basins ($\mathcal{B}$). Based on the SLOSH model, the National Hurricane Center (NHC) developed a SLOSH Display Package (SDP) application~\cite{SDP}, which allows users to create storm surge scenarios in each $\mathcal{B}$ by specifying several of the hurricane parameters. In this study, we utilize Maximum Envelope of Water (MEOW) data generated from SDP to assess storm surge scenarios and identify substations at risk of flooding. The MEOW dataset consists of the maximum recorded surge depth at each grid location within a basin for a given hurricane category, translational speed, and direction of motion but with varying landfall points. If the MEOW value obtained from SDP is $D_{\mathcal{S}}$ at a substation ${\mathcal{S}}$ located at a geographical coordinate $\mathcal{X}_{\mathcal{S}}$ for a specific ${\mathcal{B}}$, then the inundation level of ${\mathcal{S}}$ within its 0.5 mile radius is referred to as $\mathcal{X}^\mathcal{B}_{{\mathcal{S}}, D_{\mathcal{S}}}$.

\subsection{Power Systems Impact Model}
Fragility curves are proposed to assess the impact of extreme weather events on the power grid~\cite{serrano2023comprehensive}. The component fragility models provide the probability of failure of a component as a function of hazard intensity. In this work, we use the fragility of line segments, as a combined fragility of line and tower, as a function of $v_{max}$ experienced by the line segment in a power grid. The value of $v^t_{max}$ experienced by each line segment is obtained by a method discussed in~\cite{poudyal2023spatiotemporal}. The structure and strength of lines and towers differ by voltage level in a transmission grid. Hence, we modify the fragility of each line segment based on its voltage level, $\mathcal{V}$.  

\begin{equation}
\small
\begin{aligned}
&\mathbb{P}_{out}^{t,\zeta}(l_\mathcal{V})= \begin{dcases} 0 & \Gamma_{l_\mathcal{V}}^{t,\zeta} < v^{l_\mathcal{V}}_{cri} \\
\frac{\Gamma_{l_\mathcal{V}}^{t,\zeta} - v^{l_\mathcal{V}}_{cri}}{v^{l_\mathcal{V}}_{col} - v^{l_\mathcal{V}}_{cri}}  & v^{l_\mathcal{V}}_{cri} \leq \Gamma_{l_\mathcal{V}}^{t,\zeta} < v^{l_\mathcal{V}}_{col} \\
1 & \Gamma_{l_\mathcal{V}}^{t,\zeta} \geq v^{l_\mathcal{V}}_{col} \end{dcases} \\
\end{aligned}
\label{eq:hurricane_outage}
\end{equation}
\noindent
where, 
$\mathbb{P}_{out}^{t,\zeta}(l_\mathcal{V})$ is the outage probability of line $l$ with voltage level $\mathcal{V}$ for hurricane $\zeta$ at time step $t$, and $\Gamma_{l_\mathcal{V}}^{t,\zeta}$ is the maximum sustained wind speed experienced by $l_{\mathcal{V}}$ due to $\zeta$ at $t$. Here, $v^{l_\mathcal{V}}_{cri}$ is the wind speed beyond which a branch is affected, and $v^{l_\mathcal{V}}_{col}$ is the wind speed above which a branch collapses. The values of $v^{l_\mathcal{V}}_{cri}$ and $v^{l_\mathcal{V}}_{col}$ differ for each $l_\mathcal{V}$.

The substation failure probability depends on the corresponding inundation level. The storm surge impact on the substation is determined using the following Weibull stretched exponential function~\cite{poudyal2023spatiotemporal}.

\begin{equation} 
\small
\mathbb{P}^{\mathcal{S}}_{out}(\mathcal{X}^\mathcal{B}_{\mathcal{S}, D_{\mathcal{S}}}) = 1 - exp\left[- {\left(\frac{\mathcal{X}^\mathcal{B}_{\mathcal{S}, D_{\mathcal{S}}}}{a}\right)^b}\right] \\
\label{eq:flood_outage}
\end{equation}

\noindent
where $\mathbb{P}^{\mathcal{S}}_{out}(\mathcal{X}^\mathcal{B}_{\mathcal{S}, D_{\mathcal{S}}})$ is the probability of experiencing the inundation level around 0.5 miles of $\mathcal{S}$ having inundation of $D_{\mathcal{S}}$ for any $\mathcal{B}$. The fragility model includes two constants $a\in \mathbb{R}^+$ and $b>2$ with known values. If a substation experiences an outage due to flooding, the transmission lines linked to that substation, inward or outward, are also considered out-of-service for the rest of the storm period. Let $\mathcal{L}^t_{Total} = \mathcal{L}^t_\mathcal{W} \cup \mathcal{L}^t_\mathcal{F}$ be a set of lines that are out of service due to the combined effect of hurricane wind and flood damage at every time step $t$. Here, $\mathcal{L}^t_\mathcal{W}$ and $\mathcal{L}^t_\mathcal{F}$ refer to the set of outage lines due to hurricane wind damage and flood-induced substation outages, respectively.

Due to the disruption of branches, several buses are islanded from the grid either without generators (offline buses) or with generators that can fully or partially meet the demand of the islanded grid. Islands with more generation than load require generation curtailment, while those with loads higher than generation require load curtailment. The load shed from each bus is based on priorities set by the operators on load criticality. If information on load criticality is missing, then the load in each bus will be curtailed by a uniform percentage, reflected by the load deficit relative to the total load demand on the corresponding island. 

\vspace{-0.1cm}
\subsection{Community Impact Assessment}
In order to assess the impact on the community, it is essential to know the geographical distribution of load losses due to hurricanes. The synthetic test cases identify the buses and substations based on zip codes~\cite{syntheticgrid_Adam}. This work correlates bus load loss with zip code locations. Multiple substations may exist in cities with greater load density. The losses at the zip code level are aggregated into city load loss by summing the losses from all zip codes belonging to the same city. Furthermore, a county-level analysis is proposed to alleviate the concern of loads being served from the buses in one city to the neighboring cities.
After identifying cities and their associated load loss, cities within the same county, as indicated by United States Postal Service data~\cite{simplemaps}, are aggregated to find the county-level load loss. Cities that reside in multiple counties have their load distributed based on the percentage of the city's population residing in each county. Let $\mathcal{P}^D_\mathcal{C}$ be the total load demand in each county ($\mathcal{C}$) and the load loss in $\mathcal{C}$ at each time step $t$ due to corresponding $\zeta$ is represented by $\hat{\mathcal{P}}^{t,\zeta}_\mathcal{C}$. If $\mathcal{T}^\zeta$ is the total time step of each $\zeta$, then the corresponding load loss associated with every $\mathcal{C}$ is obtained as, 

\begin{equation}
\small    \hat{\mathcal{P}}^{\zeta}_\mathcal{C} = max\{\hat{\mathcal{P}}^{t,\zeta}_\mathcal{C}\}~ \forall t \in \mathcal{T^\zeta}
\end{equation}

\begin{figure*}[t]
    \centering
    \includegraphics[width=1.0\linewidth]{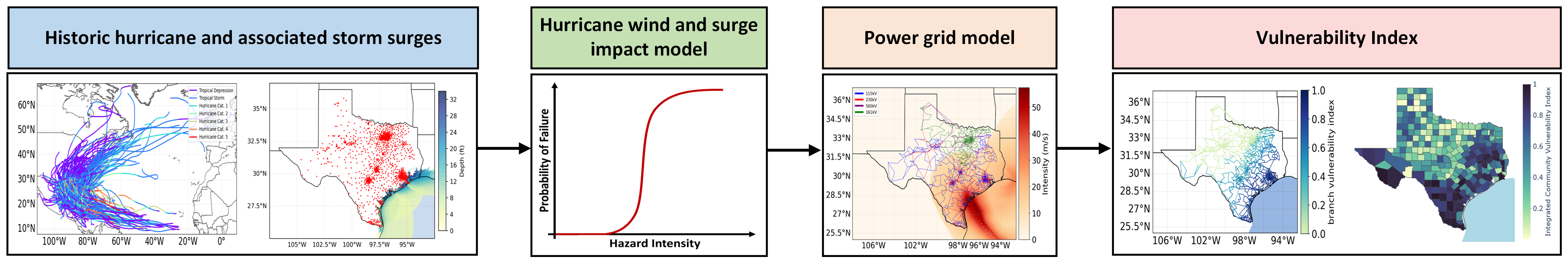}
    \caption{Overall framework of assessing hurricane and storm surge-induced power system
vulnerabilities and their socioeconomic impact}
\vspace{-0.5cm}
    \label{fig:framework}
\end{figure*}

\section{Vulnerability Assessment}

\subsection{Power System Vulnerability}
In this work, we assess the power system's vulnerability as the vulnerability of the branches and substations due to hurricanes and storm surge scenarios. Although hurricane tracks are obtained from IBTrACS, the storm surge scenarios are probabilistic, which considers several hurricane scenarios with similar characteristics. Due to a lack of surge data for hurricanes with lower intensity, we assume that such scenarios do not trigger any surge, and the impact is due to the hurricane alone. The spatiotemporal outage probabilities of branches and substations are obtained for all the hurricanes and storm surges using the model described in Section~\ref{sec:weather_impact_model}. The maximum outage probability experienced by each branch for each hurricane is observed to determine the vulnerability of these components. For substations, we observe the maximum outage probability averaged over all basins. Finally, the percentile rank is computed for each branch and substation to obtain the branch vulnerability index (BVI) and substation vulnerability index (SSVI).

\subsection{Community Vulnerability}
To observe the community vulnerability due to the impact of hurricanes and storm surges on the power grid, we obtain the value of $\hat{\mathcal{P}}^{\zeta}_\mathcal{C}$ and normalize the load loss using $\mathcal{P}^D_\mathcal{C}$ to get the vulnerability of the community in each $\mathcal{C}$ for each $\zeta$. We then define expected vulnerability of each $\mathcal{C}$ based on all $\zeta$ as $\hat{\mathcal{P}}_\mathcal{C}$ = $\mathbb{E}_\zeta\left( \frac{\hat{\mathcal{P}^{\zeta}}_\mathcal{C}}{\mathcal{P}^D_\mathcal{C}}\right)$ and term the outage vulnerability index (OVI) as
\vspace{-0.1cm}
\begin{equation}
    OVI = percentile.rank\left(\hat{\mathcal{P}}_\mathcal{C}\right)
    \label{eq:OVI}
\end{equation}

\begin{figure}[t]
    \centering
    \includegraphics[trim={16cm 2cm 10cm 2cm}, clip,width=0.7\linewidth]{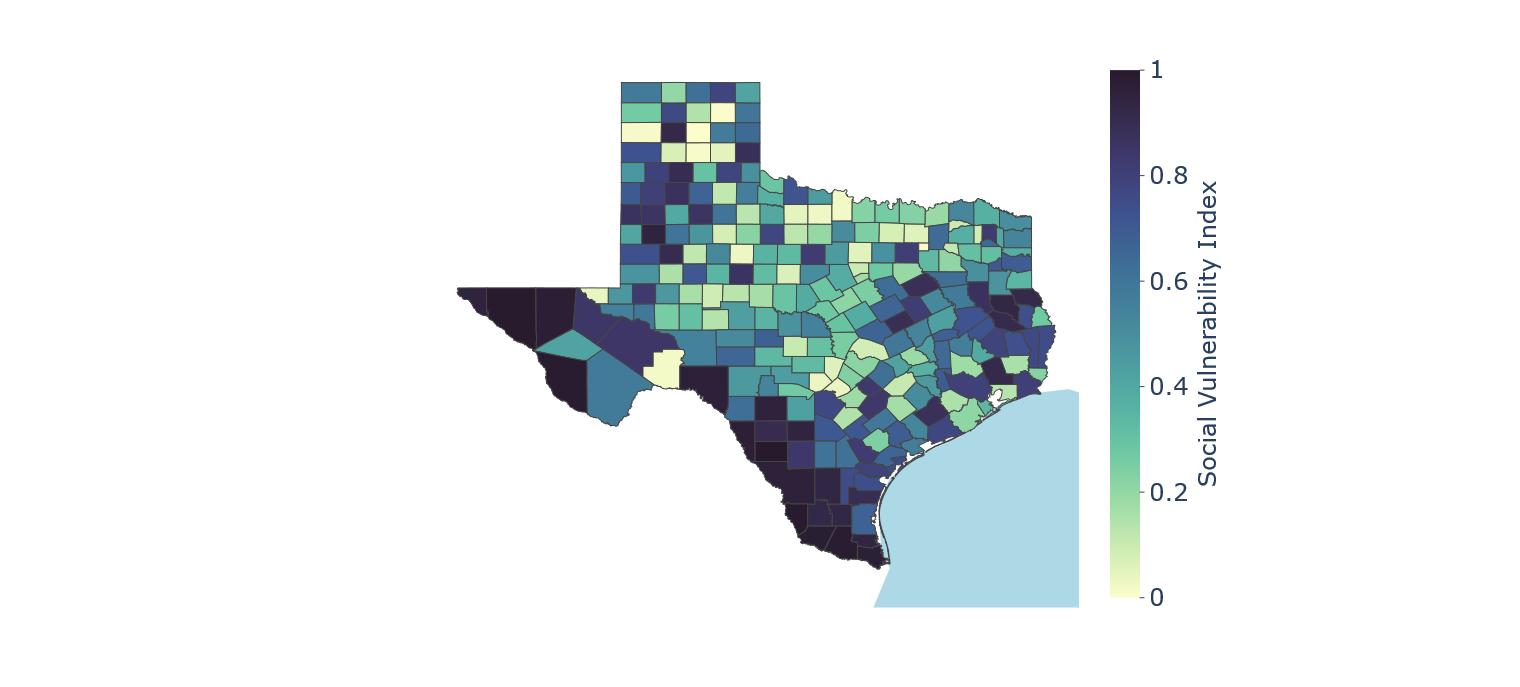}
    \caption{CDC's social vulnerability index}
    \vspace{-0.5cm}
    \label{fig:CSCSVI}
\end{figure}

Although OVI defines the propagating impact of hurricanes and storm surges on the community through their grid impact, it does not incorporate a socioeconomic impact. To address this concern, we leverage the Centers for Disease Control and Prevention (CDC)’s Social Vulnerability Index (SVI) for socioeconomic analysis~\cite{SVI}, see Fig.~\ref{fig:CSCSVI}. The SVI metric aggregates vulnerability based on four major themes: socioeconomic status, racial/ethnic minority, household characteristics, and housing/transportation type. These themes are then normalized and ranked by percentile into an encompassing index, SVI, that ranges between 0 and 1 (1 being the highest vulnerability). SVI has been widely adopted in relating infrastructure losses due to HILP events to their associated socioeconomic vulnerabilities~\cite{Soc_Tech}. Further documentation on SVI can be found at~\cite{SVI}. Each of SVI's four themes are comprised of socioeconomic variables in sum of percentile $\mathbb{S}_i$, where $i$ denotes each theme. All four themes are summed together to create $\mathbb{S}$, then finally percentile-ranked again to achieve the Social Vulnerability Index seen in (\ref{eq:SVI}). 

\vspace{-0.1cm}
\begin{equation} 
\small
SVI=percentile.rank(\sum_{i=1}^{4} \mathbb{S}_i)
\label{eq:SVI}
\end{equation}

The social and outage vulnerability metrics are compounded to create a new measurement, named Integrated Community Vulnerability Index (ICVI), which is a modified SVI that accounts for power system losses. The percentage of county population without power is treated as a 5th theme to SVI; therefore, it is min-max normalized and added to the min-max normalized $\mathbb{S}_i$, then percentile ranked between 0 and 1. This adjusts the original SVI with sensitivity to the outage analysis obtained.

\vspace{-0.5cm}
\begin{equation} 
\small
\begin{gathered}
ICVI=percentile.rank(\lVert\sum_{i=1}^{4}  {\mathbb{S}_i}\rVert + \lVert\hat{\mathcal{P}}_\mathcal{C}\rVert)
\end{gathered}
\label{eq:ICVI}
\end{equation}

Finally, with (\ref{eq:ICVI}), we can simultaneously analyze the socioeconomic and power system impacts. 

\section{Simulation Framework}
To observe the overall vulnerability due to the impact of hurricanes and storm surges on the power grid, the parameters for historical hurricanes are obtained from IBTrACS, and storm surge scenarios are generated based on the hurricane parameters in SDP. The time-varying outage probability for each of the power system components is obtained using (\ref{eq:hurricane_outage}) and (\ref{eq:flood_outage}). Several MCSs are conducted to obtain the $\hat{\mathcal{P}}^\zeta_\mathcal{C}$ based on the methods described in Section~\ref{sec:weather_impact_model}. Finally, the community-level vulnerability due to the impact of extreme weather events on the power grid is quantified as ICVI, using (\ref{eq:ICVI}). This section details the overall simulation framework, as shown in Fig.~\ref{fig:framework}.

Initially, historical hurricanes that made landfall in the Texas coastal area from 2000 to 2018 were obtained using IBTrACS. There were a few hurricanes that did not affect the region of interest. Hence, we set a boundary region and filter the hurricanes that impact the boundary region at any point during its occurrence. Within the investigation region, 97 storms were identified within the Gulf of Mexico region, of which 26 showed activity within the boundary region. Furthermore, there were a few hurricanes with missing information on $R_{v_{max}}$ and $R_s$. In such cases, the parameters were estimated from their kernel density estimates as described in~\cite{9917119}. As per NHC's guidelines, all storms are labeled from category -1 to 5 on the Saffir-Simpson hurricane wind scale, where categories -1 and 0 refer to tropical depression and tropical storm, respectively. 
SDP can only provide the storm surge information for category 1 and higher. Hence, we identify 11 storms out of 26 that trigger storm surges on the coast and are considered to have an inundation impact in this study. MEOW data are generated from SDP for each of these 11 historic hurricanes at five basins of Texas --- namely Corpus, Galveston, Laguna, Matagorda, and Sabine. Each basin's inundation level depends on the category, landfall direction, hurricane's translational speed, and mean to high tide conditions.  

The wind and storm surge scenarios are then integrated into the geographical footprint of Texas with a synthetic ERCOT 2000-bus system~\cite{syntheticgrid_Adam}. The power grid model comprises 1250 substations and 1918 transmission lines operating at different voltage levels: 115kV, 161kV, 230kV, and 500kV, with a total load of 67.11 GW and a generation capacity of 96.29 GW. The load substations are clustered based on geographical and average load consumption, and the buses have additional information on the zip code or the city name in Texas, which we leverage to map the load associated with each county. The value of $\hat{\mathcal{P}}^\zeta_\mathcal{C}$ is computed using MCS conducted on each $\zeta$ and $\mathcal{B}$. As discussed, only 11 out of 26 hurricanes instigate storm surges. Hence, we only consider obtaining $\mathcal{L}^t_\mathcal{W}$ for the remaining 15 hurricanes. Since storm surges are static models, we use a surge activation flag that gets triggered once the hurricane is within 6 hours from landfall. We only consider the impact of storm surge beyond this time frame for any $\zeta$ with additional surge impact. It is assumed that substations in the coastal regions are elevated at a height of 3 meters as an additional planning measure. MCS provides several outage scenarios corresponding to branch and substation outages. For each MC scenario, DC optimal power flow is conducted to obtain the load loss associated with that scenario at each $t$, and the loss is then mapped back to the county using the method described in Section~\ref{sec:weather_impact_model}. The community-level vulnerability is quantified as ICVI. Initially, OVI is calculated to assess the outage vulnerability due to hurricanes and storm surges using (\ref{eq:OVI}). We then blend outage likelihood with SVI to compute ICVI based on (\ref{eq:ICVI}).
\section{Results and Analysis}
In this work, the hurricanes are extracted from IBTRaCS and analyzed using CLIMate ADAptation (CLIMADA) package~\cite{Climada}. The synthetic 2000-bus ERCOT system~\cite{syntheticgrid_Adam} is modeled in MATPOWER, which is further utilized for DC power flow. 
The fragility model discussed in (\ref{eq:hurricane_outage}) is based on the voltage level of the branches. Table.~\ref{tab:fragility_data} presents the values of $v^{l_\mathcal{V}}_{cri}$ and $v^{l_\mathcal{V}}_{col}$ used in this work. These values are arbitrarily selected for simulation and can be modified based on data availability.

\begin{table}[!ht!]
    \centering
    \caption{voltage level-based critical and collapse sustained wind speed values for each line segment for fragility analysis}    
    \begin{tabular}{c|c|c}
    \hline
         Voltage level (kV) & $v^{l_\mathcal{V}}_{cri}$ (m/s) & $v^{l_\mathcal{V}}_{col}$ (m/s)  \\[1ex]
    \hhline{===}
        115 & 25 & 55   \\
    \hline
        161 & 30 & 60   \\
    \hline
        230 & 35 & 65   \\
    \hline
        500 & 45 & 75  \\
    \hline
    \end{tabular}
    \vspace{-0.2cm}
    \label{tab:fragility_data}
\end{table}

\begin{figure}[t]
    \subfigure[]{
    \centering
    \includegraphics[width=0.485\linewidth]{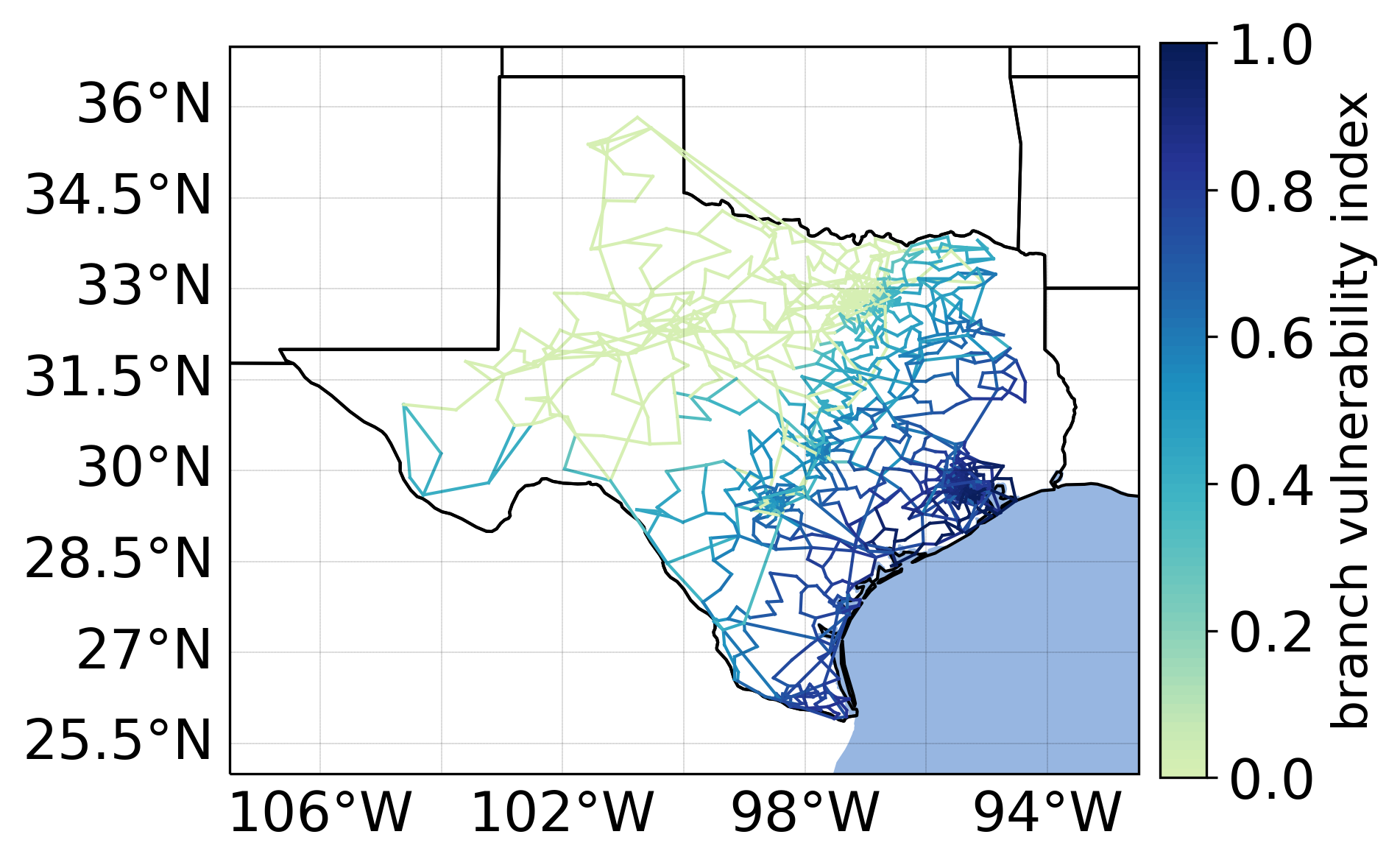}}
    \hspace*{-1em}
    \subfigure[]{
    \centering
    \includegraphics[width=0.485\linewidth]{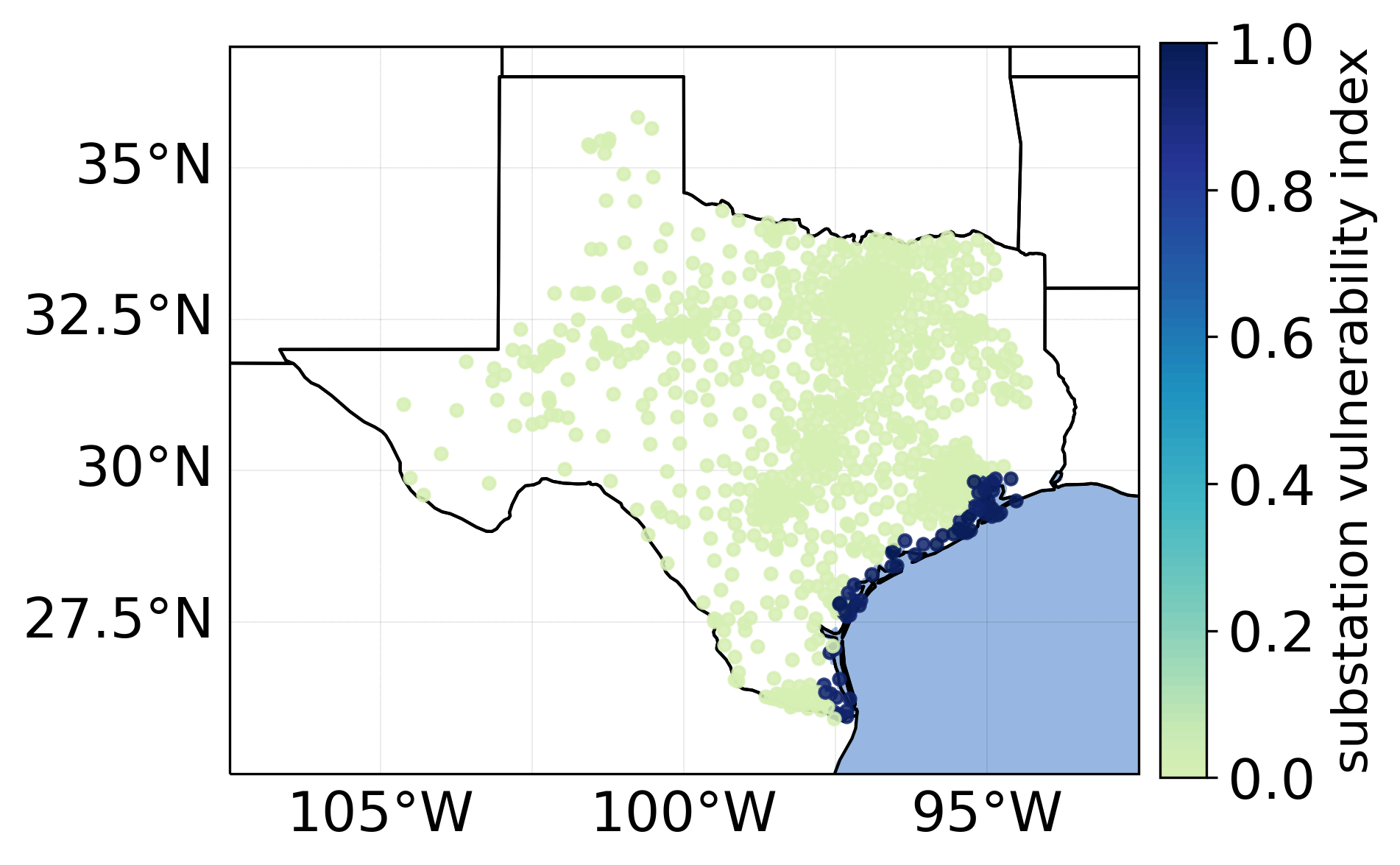}}    
    \caption{Vulnerability indices based on the percentile rank of the average outage probability of a) branches and b) substations for all hurricanes.}
    \vspace{-0.5cm}\label{fig:branch_sub_vulnerability}
\end{figure}

Fig.~\ref{fig:branch_sub_vulnerability} represents the branch and substation vulnerability index for all $\zeta$ and $\mathcal{B}$. Since hurricanes and storm surges have the highest impact on the coastlines, the most vulnerable components lie around the coastal regions. The vulnerability of inner coastal regions depends on the fragility of the branches. It is to be noted that branch and substation vulnerabilities, in Fig.~\ref{fig:branch_sub_vulnerability}, are based on outage probability rather than the load loss they incur in the system if they are damaged.

\begin{figure}[t]
    \centering
    \includegraphics[width=0.6\linewidth]{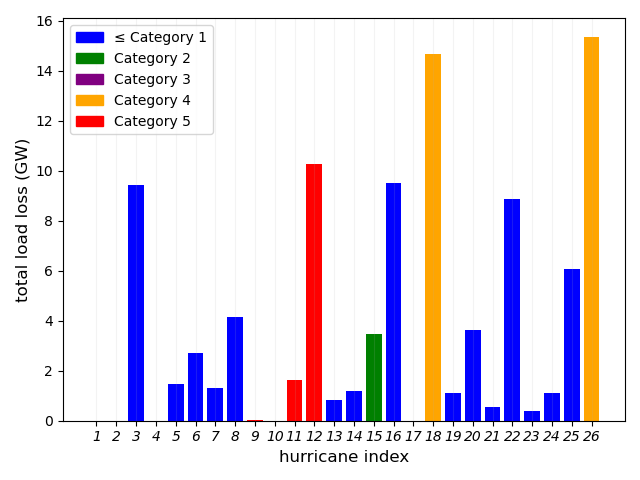}
    \caption{Bar chart depicting an overall load loss in the study area caused by historical hurricanes}
    \label{fig:hurricane_loss_barplot}
\end{figure}

Fig.~\ref{fig:hurricane_loss_barplot} shows a bar chart illustrating the overall load loss across the system attributed to historical hurricanes with their respective category. It can be seen that hurricanes Rita, Ike, and Harvey, with respective indices of 12, 4, and 26, exhibit substantial impact on the system, resulting in losses of 10.2 GW, 14.67 GW, and 15.2 GW. These are the three most formidable hurricane events ever recorded in Texas~\cite{disaster}. Interestingly, it is observed that some hurricanes with lower categories exhibit more significant effects, such as Hurricane Allison (index=3), than those with higher categories, such as Hurricane Ivan (index=9). This discrepancy is attributed to the fact that hurricane Allison, despite its 0 category, generates heavy flooding over coastal Texas. In contrast, by the time Hurricane Ivan reached Texas, it had significantly weakened, leading to a reduced load loss, while its substantial impact was observed in a location other than Texas.

\begin{figure}[]
    \subfigure[]{
    \centering
    \includegraphics[trim={16cm 3cm 10cm 2cm}, clip,width=0.485\linewidth]{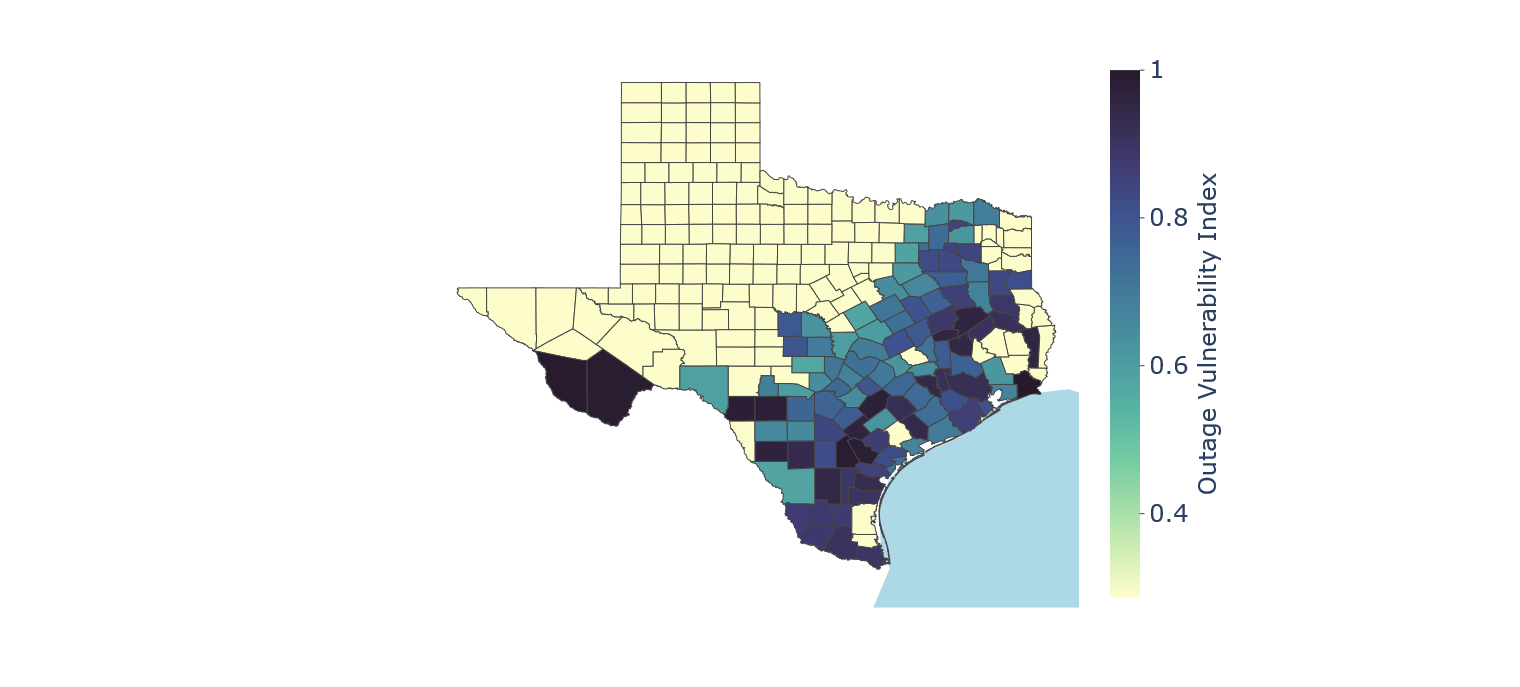}
    }
    \hspace*{-1.5em}
    \subfigure[]{
    \centering
    \includegraphics[trim={16cm 3cm 10cm 2cm}, clip,width=0.485\linewidth]{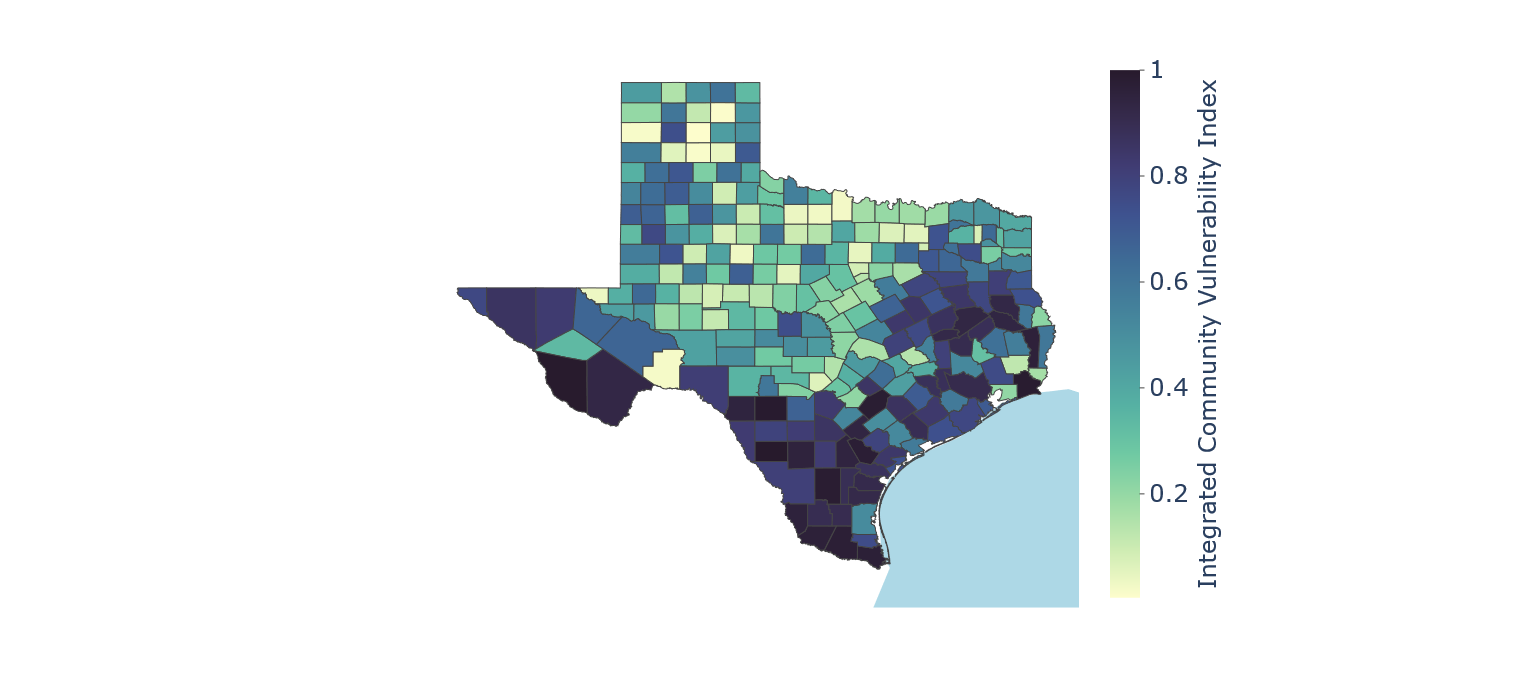}
    }    
    \caption{a) Outage Vulnerability Index (OVI) and b) Integrated Community Vulnerability Index (ICVI) due to historical hurricanes in the ERCOT region.}
    \label{fig:OVI_ICVI}
\end{figure}

Fig.~\ref{fig:OVI_ICVI}a shows the OVI obtained using (\ref{eq:OVI}). It is evident that communities with higher outage vulnerability from hurricanes and storm surges are predominantly situated along the coastline. However, because of the grid's architecture, a hurricane-induced outage can still be affected by communities farther from the coastline, as seen in Fig.~\ref{fig:OVI_ICVI}. The compounded effect of their social vulnerability amplifies this impact, highlighting the need for infrastructure improvement in these regions. This can be useful knowledge for transmission planners when determining vulnerable components to future climate events. In Fig.~\ref{fig:OVI_ICVI}b, the ICVI can be compared to results of SVI in Fig.~\ref{fig:CSCSVI} and OVI in Fig.~\ref{fig:OVI_ICVI}a. ICVI reflects the socioeconomic disparities from SVI and highlights communities disproportionately impacted by hurricane and storm surge-induced outage events. There is a shift in vulnerability towards the coastline in ICVI compared to SVI. This gives important insights to grid operators on allocating resources during future hurricane events because we know communities of different vulnerabilities respond differently to outages. A limitation in this analysis is that some counties with very low populations do not have outage data as they are served from different regions; also, counties not within ERCOT territory are not analyzed. However, most of these counties are in regions further from the coastline and less likely to be affected.

\section{Conclusion}

This paper presents a probabilistic evaluation of the combined effects of historical hurricanes and storm surges on the synthetic power grid. It also introduces a community vulnerability metric that integrates the CDC’s SVI with the outage factor derived from the results of the impact assessment. Simulation results show that a hurricane of a higher category does not necessarily entail higher impacts on the system; it is also influenced by the landfall location and associated storm surge. The communities farther from the coast are not immediately perceived as high-risk due to their distance from the coast and often get less attention in storm management strategies. However, the community could have higher socioeconomic vulnerability exacerbated by the outage vulnerability. Hence, ICVI provides a holistic assessment of vulnerable communities by incorporating the effect of extreme weather events and their impact on power outages and associated communities. The proposed methodology can assist planners in identifying the vulnerable system components and vulnerable communities, enabling strategic planning to minimize outages or expedite the restoration while considering energy equity as an additional dimension. In the future, we aim to observe the impact of changing climate on the extreme event scenarios, how other HILP events impact disadvantaged communities, and evaluate the underlying assumption that load and population are highly correlated.


\bibliographystyle{IEEEtran}
\bibliography{ref.bib}

\end{document}